# Crosstalk Noise based Configurable Computing: A New Paradigm for Digital Electronics

Naveen Kumar Macha, Md Arif Iqbal, Bhavana Tejaswini Repalle, Sehtab Hossain, Mostafizur Rahman
Computer Science Electrical Engineering, University of Missouri-Kansas City, MO, USA
nmhw9@mail.umkc.edu, mibn8@mail.umkc.edu, rahmanmo@umkc.edu

*Abstract*—The past few decades have seen exponential growth in capabilities of digital electronics primarily due to the ability to scale Integrated Circuits (ICs) to smaller dimensions while attaining power and performance benefits. That scalability is now being challenged due to the lack of scaled transistor performance and also manufacturing complexities [1]-[5]. In addition, the growing cyber threat in fabless manufacturing era poses a new front that modern ICs need to withstand. We present a new noise based computing where the interconnect interference between nanoscale metal lines is intentionally engineered to exhibit programmable Boolean logic behavior. The reliance on just coupling between metal lines and not on transistors for computing, and the programmability are the foundations for better scalability, and security by obscurity. Here, we show experimental evidence of a functioning Crosstalk computing chip at 65nm technology. Our demonstration of computing constructs, gate level configurability and utilization of foundry processes show feasibility. These results in conjunction with our simulation results at 7nm for various benchmarks, which show over 48%, 57%, and 10% density, power and performance respectively, gains over equivalent CMOS in the best case, show potentials. The benefits of Crosstalk circuits and inherent programmable features set it apart and make it a promising prospect for future electronics.

## I. INTRODUCTION

Traditionally, the inference between interconnects is considered a curse. This interference is more prominent in nanoscale ICs and in lower metal lines. With Crosstalk computing, we aim to turn this curse into a feature. We intentionally arrange metal lines such that they can interfere in a deterministic manner. Then we capture this deterministic interference in a certain timeframe to ascertain logic. Let us use an example of a two input (A and B) logic. In Crosstalk, we would drive these inputs in two adjacent metal lines, and in between those lines, we will have another metal line to capture the interference (or the output). In interconnect terminology, the driving inputs would be called Aggressors, and the interference capturing line would be called the Victim. For capturing interference, the Victim would be intentionally kept floating (not connected to power supply or ground). As the Aggressors transition from $0 \rightarrow 1$ or $1 \rightarrow 0$, corresponding interference would result in voltage gain or drop in the Victim node. If any of the input transitions (A or B) from 0 to 1, results in a sufficiently high voltage induction in Victim, we would achieve OR logic, and if only when both A and B transitions from 0 to 1, we notice high voltage induction in Victim node, we would call the metal arrangement as performing AND logic.

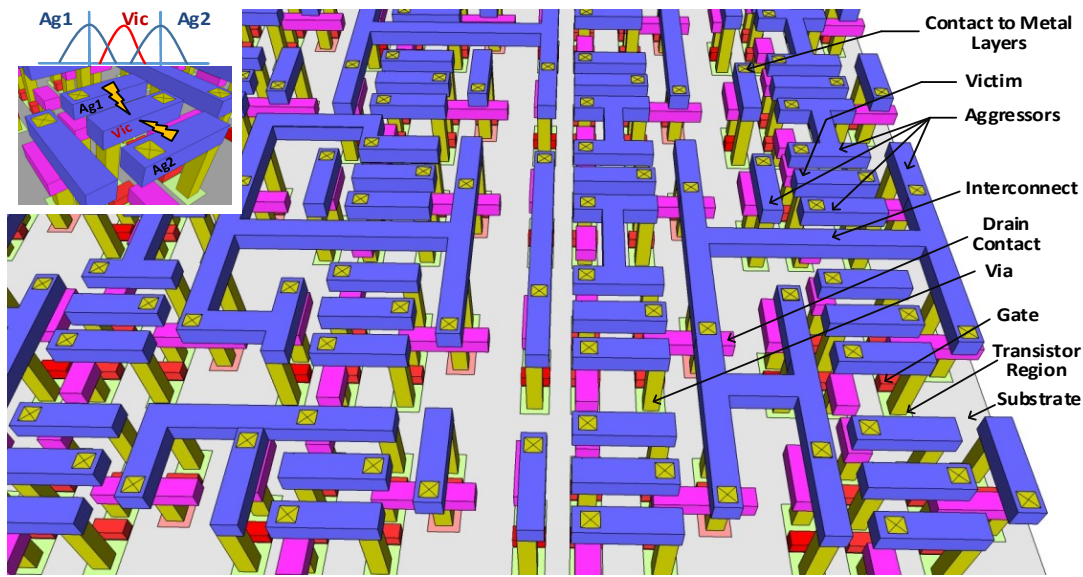

**Fig. 1.** Abstract view of the Crosstalk computing fabric.

The idea is illustrated in Fig. 1, where the metal lines are arranged on the top and the controlling transistors are at the bottom. A control transistor gated by a Discharge clock is required (Fig. 2a.i) to keep the Victim node floating (ready for interference induction) and the inverter attached to the Victim node is required to achieve complete voltage swing for next stages. Fundamentally though, the signal induction happens without the help of transistor, and is a key distinguishing factor from CMOS, where PMOS and NMOS transistors are arranged and gated in a complementary manner to achieve logic function. The less reliance on transistors are also foundations for higher density and gains on other metrics. For example, a 2-input NAND gate in Crosstalk computing requires only 3 transistors, whereas

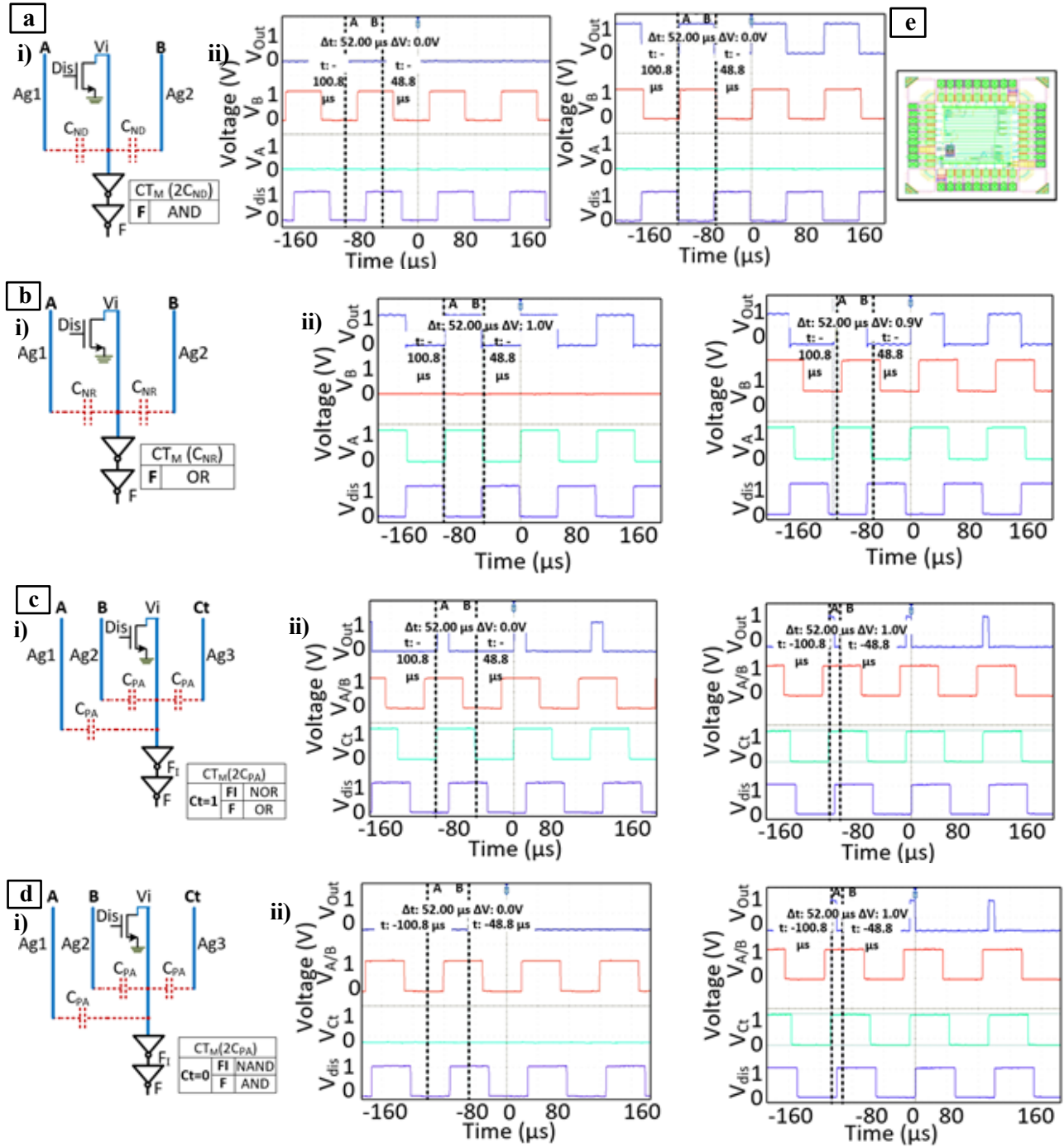

**Fig 2**. Experimental results of Crosstalk gates. a) Crosstalk AND gate, b) Crosstalk OR gate, c) Reconfigurable Crosstalk OR gate (Ct=1), d) Reconfigurable Crosstalk AND gate (Ct=0), e) Layout of fabricated Crosstalk chip (details in Supplementary Doc)

CMOS requires 4, which is suggestive of over 2x density benefits.

## II. FOUNDATIONAL LOGIC GATES

The schematic for Crosstalk AND and OR gates along with their experimental measurement results are shown in Fig 2.a&b. Notably, both AND and OR gates have the exact same configuration with only difference of coupling capacitance. The coupling capacitance dictates the AND or OR behavior by having less or more strength. For the case of Fig. 2b, the $C_{NR} > C_{ND}$, which ensures that any of the signals A or B transitioning, resulting in logic1. The coupling capacitance is a distinguishing factor and can be engineered by tuning dielectric parameters or metal dimensions during fabrication. The less reliance on device also implies better scalability, which can be determined by circuit integration and scaling/arrangements (e.g., 2-D, 3-D) of metal lines. Since, any function can be decomposed into NAND-NAND, NOR-NOR, AND-OR, we can conclude that any logic function can be implemented with Crosstalk computing – hence this computing approach is functionally complete and universal. For large-scale circuits, logic cascading and maintaining signal integrity is a critical issue. In this regard, the Crosstalk computing approach provides opportunities as well as challenges. Since utilizing Crosstalk, we can implement both fundamental logic gates (AND, NAND, XOR, OR, NOR) and also reduce complex combinational logic blocks, any logic function can be implemented. Key requirements for cascading and implementing large circuits are drive strength of outputs and noise margin. To achieve these, we incorporate CMOS inverters and buffers which are logically transparent but electrically behave as voltage or signal boosters because of their static connection to power rails. More importantly, their drive-strength can be increased in proportion to the fan-out load (i.e., number of gates connected to the output), by increasing current drawn from the power rails. To maintain synchronization between cascaded logic stages the discharge transistor that is connected to floating victim nodes are controlled in alternative clock cycles. Detail discussion on Crosstalk circuit design methodology and signal integrity issues are presented in supplementary section.

## III. PROGRAMMABLE LOGIC GATES

A unique feature of Crosstalk computing is the programmability. Crosstalk gates can be configured at run-time. Since Crosstalk circuits are identical (e.g., 2-input AND vs. 2-input OR) with the only difference of the coupling capacitance, if the capacitance can be tuned different logic can be achieved. We achieve this tuning by introducing a new Aggressor and call it the control input (Ct). If the control input is turned ON, it is as if an extra coupling strength/bias is introduced and as a result the circuit behaves differently. For example Fig. 2.(c&d).i shows the Crosstalk programmable AND2-OR2 circuit schematic. The inputs A, B has the same coupling $C_{PA}$. Ct aggressor receives $2C_{PA}$ capacitance. A table adjacent to the circuit diagram lists the margin function and the circuit operating modes. The margin function for AND2-OR2 cell is $CT_M (2C_{PA})$, which makes it behave as AND2 gate when control Ct=0. Whereas, when Ct=1, it infers an extra charge through coupling capacitance $2C_{PA}$ and effectively manipulates the margin function to $CT_M (C_{PA})$, making it an OR2 gate. Following the function $CT_M (C_{PA})$, the transition of either A or B is now sufficient to flip the inverter; thus, the gate would be now biased to operate as OR2 gate. The same response can be observed in the simulation response plots as shown in Fig. 2.(c&d).ii. The first graph shows the output F, second graph shows input A/B and third graph shows control signal and fourth panel shows dis signal. Details of measurement techniques along with all the combination are discussed in supplementary section. It can be observed that the circuit responds as AND2 when Ct=0 for input combinations (10 to 11), whereas, it responds as OR2 when Ct=1 during next eight combinations (10 to 11).

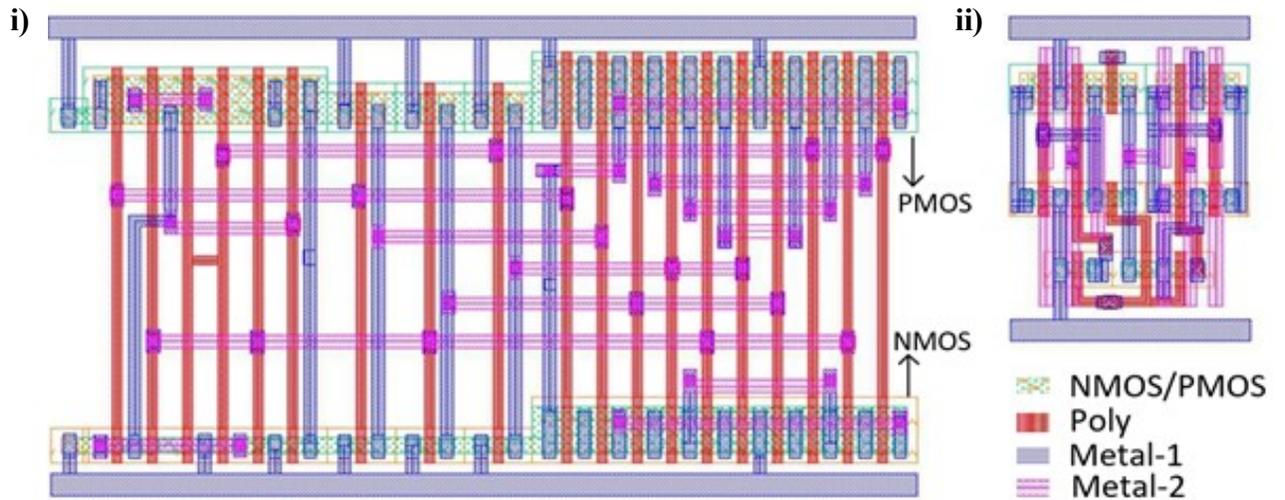

**Fig. 3.** Layouts of Full Adder Circuit (Sum and Carry): i) CMOS Layout ii) Crosstalk Layout (8)

The key benefit of this type of configurability is run-time programmability. It can be the foundation for anti-counterfeiting, resource sharing and fault-tolerant computing. In the case of anti-counterfeiting, the reconfigurability would allow cloaking/camouflaging of functionality. In our previous work we have shown a wide range of polymorphic logic behavior between following functions, AND2-OR2, AND3-OR3, AO21-OA21, AND-AO21, AND3- OA21, OR3-AO21, and OR3-OA21; also, a cascaded circuit example of Adder-Multiplier-Sorter. All of these Crosstalk Polymorphic gates are uniform and modular in structure, and thus generic to scale to larger polymorphic digital systems [6]. It was shown in [7] that dynamically configurable system is the hardest to hack. In addition, the regular layout of Crosstalk logics also add a layer of security because they would be identical and difficult to trace during reverse engineering.

The polymorphism can also open up a new front for resource sharing and fault-tolerant computing, since if a portion of circuit can be configured to do the work of other portion [8]-[10]. Fig. 2c-d are examples of resource sharing also. If a series of AOI blocks are placed in a Crosstalk ALU, it can perform both the operations of OR and AND gate, needed in a microprocessor.

IV. INTUITION FOR IMPROVEMENTS AND SIMULATION RESULTS

Intuitively we can gather the merits of Crosstalk technology by inspecting the layouts. Fig 3 shows the layout of Full-Adder. For the full adder circuit, CMOS implementation requires 40 transistors in cascaded topology (12 transistors for each XOR gates and 12 for carry logic), whereas the Crosstalk implementation requires just 13 and the interconnection requirements are also considerably less. It is evident from Fig.3 that Crosstalk circuits consume less active device area compared to CMOS.

To further understand the possibilities, we have done extensive comparison between CMOS and Crosstalk [11]. We have implemented three MCNC benchmark circuits and compared density, power and performance results with respect to CMOS at 7nm. As can be seen from Fig. 4, In terms of transistor count, the highest reduction was for the mux circuit, it was 62%. For cm85a and pcle circuits the reduction in transistor count is 59% and 23% respectively. Crosstalk circuits show on average 58% power benefits over CMOS counterparts. The benefits are primarily due to the reduction in transistor count. However, the reduction in average power for the mux circuit is not much even though transistor count reduction is maximum compared to other circuits. This is because mux circuit implementation requires many pass-gate type circuit styles circuit which results in more switching activities, hence, less power reduction. On the contrary, for pcle circuit, power reduction is more because it requires less number of buffer and pass-gate type circuit styles that means less switching activity. All the implementation are given in the supplementary section. However, for cm85a and pcle circuits, Crosstalk circuits have 10% and 53% improvement in performance respectively.

To validate the work further, we have also done scalability analysis on Crosstalk basic gates. Fig 5 shows scalability study of Crosstalk NAND gate with respect to CMOS at 180nm, 65nm, 32nm and 7nm with process variations. From Fig. 5.i it can be seen that both CMOS and Crosstalk NAND gate show reduction in power; however, Crosstalk gates show ~42.5% more reduction in power than CMOS gates for all the technology nodes. The improvement in power for Crosstalk gates is because of less number of active devices and lower effective load due to the series connection of coupling capacitance to the inverter. Fig. 5.ii shows the performance results of Crosstalk gates with respect to CMOS for various process corners. For typical process corner, there is an average improvement of 34% in performance for Crosstalk gates compared to CMOS for all technologies. As shown in Figs. 5(i&ii), for slow process corner, the performance is the worst due to slow PMOS and NMOS devices whereas for the FF corner the performance is the best due to the fast active devices. Such performance improvement in Crosstalk circuits is due to lower effective load capacitances, lower interconnect parasitic and shorter VDD/GND to output rail.

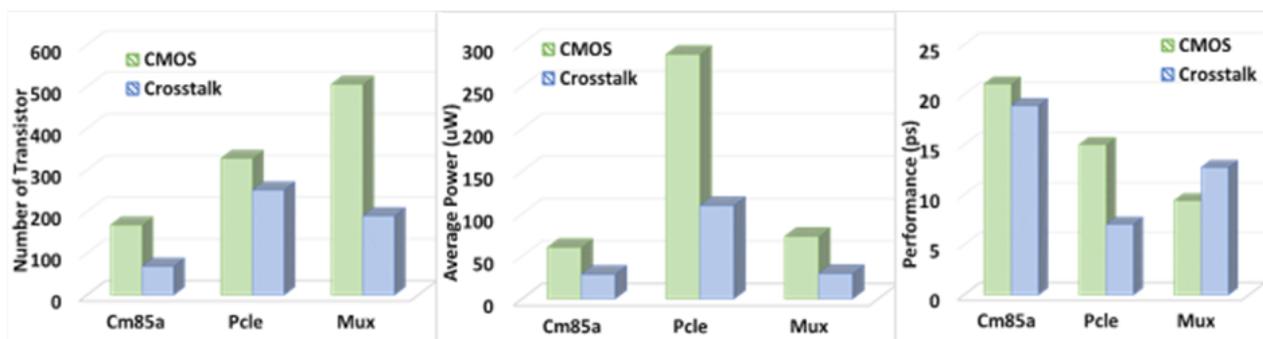

**Fig. 4.** Difference different large-scale MCNC Benchmark Circuits. i) Density Comparison, ii) Average Power and Performance

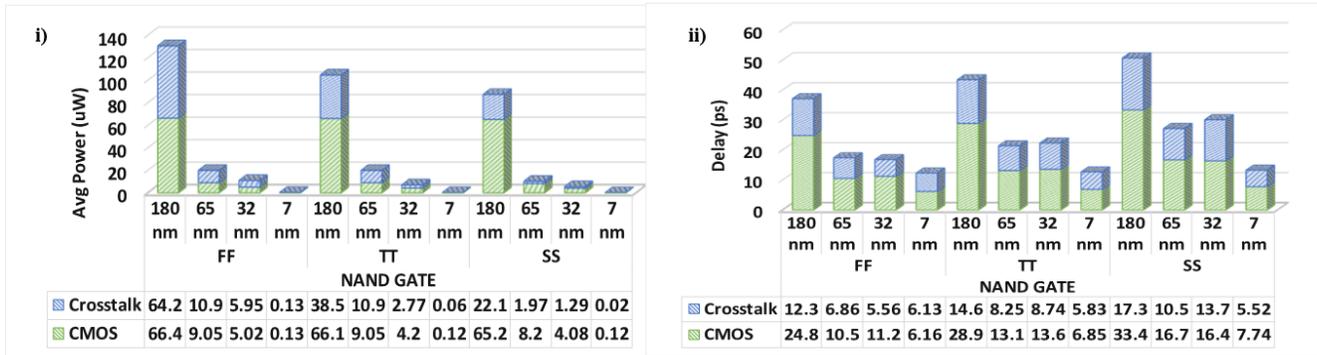

**Fig. 5.** Technology scaling impact on the performance for Crosstalk and CMOS NAND gate with process variation. i) Average power for & ii) Performance of NAND Gate at 180nm, 65nm 32nm and 7nm with process variations

## V. CONCLUSION

Crosstalk computing is a new kind of computing technique can be leveraged to build next-generation security chips. Run-time gate level reconfigurability of Crosstalk gates makes harder to hack. In this paper, we have shown excremental proof of Crosstalk computing technology at 65nm. Our experimental results have shown that the gate level configurability is feasible with existing process technique. Our results also show significant improvement in density, power, and performance for Crosstalk circuits compared to CMOS even with scaling down of technology nodes for larger-scale implementation.